\documentclass[twocolumn]{aastex631}

\begin{document}

\newcommand{\vdag}{(v)^\dagger}
\newcommand{\aastex}{AAS\TeX}
\newcommand{\latex}{La\TeX}
\newcommand{\WD}{WD~2149+021}
\newcommand{\Msun}{$M_{\odot}$}
\newcommand{\Mjup}{M$_{\mathrm{Jup}}$}
\newcommand{\Teff}{$T_{\mathrm{eff}}$}
\newcommand{\logg}{$\log{g}$}

\title{A MIRI Search for Planets and Dust Around WD 2149+021}

\author[0009-0008-7425-8609]{Sabrina Poulsen}
\affiliation{Homer L. Dodge Department of Physics and Astronomy, University of Oklahoma, 440 W. Brooks St, Norman, OK 73019, USA}

\author[0000-0002-1783-8817]{John Debes}
\affiliation{AURA for ESA, Space Telescope Science Institute, 3700 San Martin Dr, Baltimore, MD 21218, USA}

\author[0000-0002-7698-3002]{Misty Cracraft}
\affiliation{Space Telescope Science Institute, 3700 San Martin Dr, Baltimore, MD 21218, USA}

\author[0000-0001-7106-4683]{Susan E. Mullally}
\affiliation{Space Telescope Science Institute, 3700 San Martin Dr, Baltimore, MD 21218, USA}

\author[0000-0001-8362-4094]{William T. Reach}
\affiliation{Stratospheric Observatory for Infrared Astronomy, Universities Space Research Association, NASA Ames Research Center, Moffett Field, CA 94035}

\author[0000-0001-6098-2235]{Mukremin Kilic}
\affiliation{Homer L. Dodge Department of Physics and Astronomy, University of Oklahoma, 440 W. Brooks St, Norman, OK 73019, USA}

\author[0009-0004-7656-2402]{Fergal Mullally}
\affiliation{Constellation, 1310 Point Street, Baltimore, MD 21231}

\author[0000-0003-0475-9375]{Loic Albert}
\affiliation{Institut Trottier de recherche sur les exoplanètes and Département de Physique, Université de Montréal, 1375 Avenue Thérèse-Lavoie-Roux, Montréal, QC, H2V 0B3, Canada}

\author[0009-0004-6806-1675]{Katherine Thibault}
\affiliation{Institut Trottier de recherche sur les exoplanètes and Département de Physique, Université de Montréal, 1375 Avenue Thérèse-Lavoie-Roux, Montréal, QC, H2V 0B3, Canada}

\author[0000-0001-5941-2286]{J. J. Hermes}
\affiliation{Department of Astronomy, Boston University, 725 Commonwealth Avenue, Boston, MA 02215, USA}

\author[0000-0001-7139-2724]{Thomas Barclay}
\affiliation{NASA Goddard Space Flight Center, 8800 Greenbelt Road, Greenbelt, MD 20771, USA}

\author[0000-0003-1309-2904]{Elisa V. Quintana}
\affiliation{NASA Goddard Space Flight Center, 8800 Greenbelt Road, Greenbelt, MD 20771, USA}



\begin{abstract}
The launch of JWST has ushered in a new era of high precision infrared astronomy, allowing us to probe nearby white dwarfs for cold dust, exoplanets, and tidally heated exomoons. While previous searches for these exoplanets have successfully ruled out companions as small as 7-10 Jupiter masses, no instrument prior to JWST has been sensitive to the likely more common sub-Jovian mass planets around white dwarfs. In this paper, we present the first multi-band photometry (F560W, F770W, F1500W, F2100W) taken of \WD\ with the Mid-Infrared Instrument (MIRI) on JWST. After a careful search for both resolved and unresolved planets, we do not identify any compelling candidates around \WD. Our analysis indicates that we are sensitive to companions as small as $\sim$0.34 \Mjup\ outwards of 1\farcs263 (28.3 au) and $\sim$0.64 \Mjup\ at the innermost working angle (0\farcs654, 14.7 au) with 5$\sigma$ confidence, placing significant constraints on any undetected companions around this white dwarf. The results of these observations emphasize the exciting future of sub-Jovian planet detection limits by JWST, which can begin to constrain how often these planets survive their host stars evolution.
\end{abstract}



\section{Introduction} \label{sec:intro}
The majority of observed white dwarfs have hydrogen-dominated atmospheres that are expected to be nearly pure, as heavier elements gravitationally settle below the photosphere and out of our view. Despite this, 25-50\% of isolated, hydrogen-dominated white dwarfs are “polluted” with trace elements such as calcium and iron, forming the DAZ spectral class \citep{zuckerman03,koester14}. Given that these elements sink out of the atmosphere on timescales ranging from thousands of years to a few days \citep{koester09}, these metals must have recently been accreted onto the white dwarf. The origin of these accreted elements has been theorized for decades- Previously this pollution was attributed to the interstellar medium \citep{dupuis92,dupuis93,hansen03}, yet compelling evidence has accumulated which instead points to asteroids and other rocky planetesimals for the origin of these metals \citep{jura08,jura14}. While the mechanics of this are not well constrained by current observations, one widely accepted hypothesis to explain this phenomenon is the presence of giant planets in wide-orbits around the polluted white dwarf \citep{alcock86,debes02,jura03}.

In this scenario, massive planets that survive the red-giant phase occasionally perturb the orbits of asteroids and other bodies, which are then dynamically scattered towards the white dwarf. When these objects pass within the Roche limit of the star, they disintegrate into a cloud of dust and gas which then accretes onto the white dwarf. This scenario was shown to be viable by \cite{debes12}, who used numerical simulations to show that a single giant planet is capable of perturbing planetesimals into highly eccentric orbits which can then create a steady stream of material to be accreted onto the star. This model predicts that many polluted white dwarfs may host a planetary system containing a Jovian analog, although dynamical simulations have shown less massive planets are also capable of efficient scattering \citep{frewen14}.

While there are theories of second-generation planets born of mass lost on the asymptotic giant branch, white dwarfs may come to host planetary systems via first-generation planets which survive their hosts' evolution. The orbital evolution of a planet depends almost entirely on its mass and initial semi-major axis, as more massive planets in tighter orbits suffer strong tidal forces which pull the planet towards engulfment faster than the effects of stellar mass loss release the planet to wider orbits. When investigating red giants, \cite{wolthoff22} finds the global occurrence rate of planetary systems with at least one giant planet (M $>$ 0.8 \Mjup) to be 10.7\%, with a maximum occurrence rate reached at a stellar mass of 1.68 \Msun. We can also look to white dwarf progenitors (A and F stars), as we would expect the occurrence rate of giant planets in wide orbits to be similar. While the exact occurrence rates are not yet well constrained, recent results from transit and radial velocity surveys from 10 to 100 au \citep{fernandes19} show that 26\% of G and K stars between 0.1 and 100 au have a giant planet larger than 0.1 \Mjup\ and 6\% have a giant planet larger than Jupiter. Direct imaging surveys \citep{nielsen19} indicate that giant planets are even more common around A and F stars than G and K stars. For a 1.5 \Msun\ star (the estimated progenitor mass of the star discussed in this paper), \cite{mustill12} show that a 1 \Mjup\ planet contacts the stellar envelope during the AGB phase for initial circular orbits within $\sim$3.2 au. While this radius increases for larger planets, a sufficiently massive planet may survive the common envelope phase and be found interior to this limit \citep{bear12}. 

The search for giant planets around white dwarfs has been challenging due to limitations in the methods commonly used to detect exoplanets. Transit surveys, such as TESS, are designed to detect short-period objects \citep{ricker15} and are not well-suited for detecting wide-orbit planets that are of interest in this context. \citet{vanderburg2020} did succeed in using TESS to find a $\sim$13 \Mjup\ companion orbiting its white dwarf, however its orbital period is unusually short at 1.4 days and raises questions as to how this planet migrated or survived so close to its host. Radial velocity methods, which rely on detecting shifts in spectral lines, are also not ideal for white dwarfs due to their lack of prominent spectral lines for high-precision analysis. Astrometry measures the tiny movements of stars caused by the gravitational pull of orbiting planets, however Gaia is limited to closer orbits and may not be sensitive enough to detect a significant number of wide-orbit planets around white dwarfs \citep{sanderson22}. Additionally, direct detection of planets from the ground is challenging due to absorption of infrared light in the Earth's atmosphere, making it difficult to directly observe cool companions.

Investigators have historically turned to space-based searches to attempt to take advantage of the low star-planet contrast. \citet{debes05} used HST to search for companions orbiting seven of the nearest DAZs and ruled-out companions down to 10-18 \Mjup\ in orbital separations greater than 30 au. \citet{mullally07} surveyed 124 white dwarfs to look for infrared excess and placed limits on companions larger than 10 \Mjup\ less than 30 au from the star \citep[see also][]{kilic09}. \citet{brandner21} searched seven white dwarfs in the Hyades cluster with NICMOS and found no companions larger than 7 \Mjup\ beyond 10 au. There have also been successful hunts for these elusive planets- \citet{luhman12}, using Spitzer IRAC, found a cool, low-mass, brown dwarf with common proper motion to a white dwarf, while \citet{blackman21} was able to use micro-lensing to find a giant planet around a white dwarf. 

While these massive planets have been proven difficult to find in the past, JWST gives us the opportunity to look for sub-Jovian-mass companions around white dwarfs for the first time. JWST’s aperture size, coupled with its mid-infrared capabilities, provides the necessary tool to look for planets with masses $<$7 \Mjup, which are predicted to be more common and could determine whether the polluted atmospheres are truly linked to planetary systems. The cycle 1 JWST program ``A Search for the Giant Planets that Drive White Dwarf Accretion" (GO 1911: PI, S. Mullally) observed four nearby, young, polluted white dwarfs in uncrowded fields. 

The first target observed in this program was \WD, which was first noted in 1961 as a ‘white dwarf suspect’ in a paper on Lowell proper motions \citep{giclas61} where they were cataloging stars with a proper motion of more than 0.27 arcsec/year. \WD\ was listed in that paper as G93-48. In 1965, in a paper characterizing 166 white dwarfs, it was listed as a DA white dwarf \citep{eggen65}. In 1983, it was listed among the UBVRI Photometric standard stars around the celestial equator by Arlo Landolt \citep{landolt83} and as one of the Hubble UV spectrophotometric, optical spectrophotometric and optical polarimetric standard calibration objects in 1990 \citep{turnshek90} and listed as a DA3 spectral type star. Later \citet{koester05} discovered a narrow Ca K line from UVES echelle spectroscopy taken in 2002 and 2003 in the photosphere and found an abundance of [Ca/H]=$-$7.7, confirming it as an actively accreting DAZ. Further monitoring of the white dwarf has taken place, as recently as 2016 with Keck/HIRES (NASA/Keck Program N171b; PI: Redfield), where archival visible echelle spectra still show the presence of the Ca K line. This implies at least 14 years of accretion. \WD\ was observed by both the \textit{Hipparcos} and \textit{Gaia} missions, which means that any residual acceleration detected in its astrometry could reveal the presence of a substellar companion \citep{kervella19,brandt21,kervella22}. \citet{kervella22} found no significant acceleration, limiting planets with masses $<$0.7 \Mjup\ between 3-10 au.

This paper presents the limits on planet and dust detection from our observation of \WD. We describe the methods used to obtain our observations, followed by the modeling of the white dwarf's photometry based on previous observations. We then discuss our limits to detecting resolved planets, an analysis of background objects in the MIRI field of view, the limits to detecting unresolved planets, cold dust, and finally tidally heated exomoons. We conclude by discussing our primary findings, as well as discussing some considerations to keep in mind.



\section{Description of Observations} \label{sec:obs}
\WD\ was observed by MIRI in four broadband filters centered on 5.6, 7.7, 15.0, and 21.0 \micron. This multi-filter approach allows us to employ two different methods for planet detection: searching for resolved planets using direct imaging, and unresolved planets using infrared excess. At 15 \micron, most Jovian analog planets will have contrasts of $\sim$100:1, which MIRI is able to spatially resolve beyond $\sim$15 au (1.5$\times$ the FWHM of the 15 \micron\ filter). To directly image exoplanets we focus on the 15 \micron\ filter, as these observations have the longest exposure time and provide the deepest mass limits. However, we also rely on the other filters to look for evidence that a source is a background star or galaxy. Planets should be dimmer and redder than stars, and should be shaped like a point source while typical galaxies are extended. 

\begin{figure}[ht]
\vspace*{-0.5cm}
\begin{center}
    \includegraphics[width=\columnwidth]{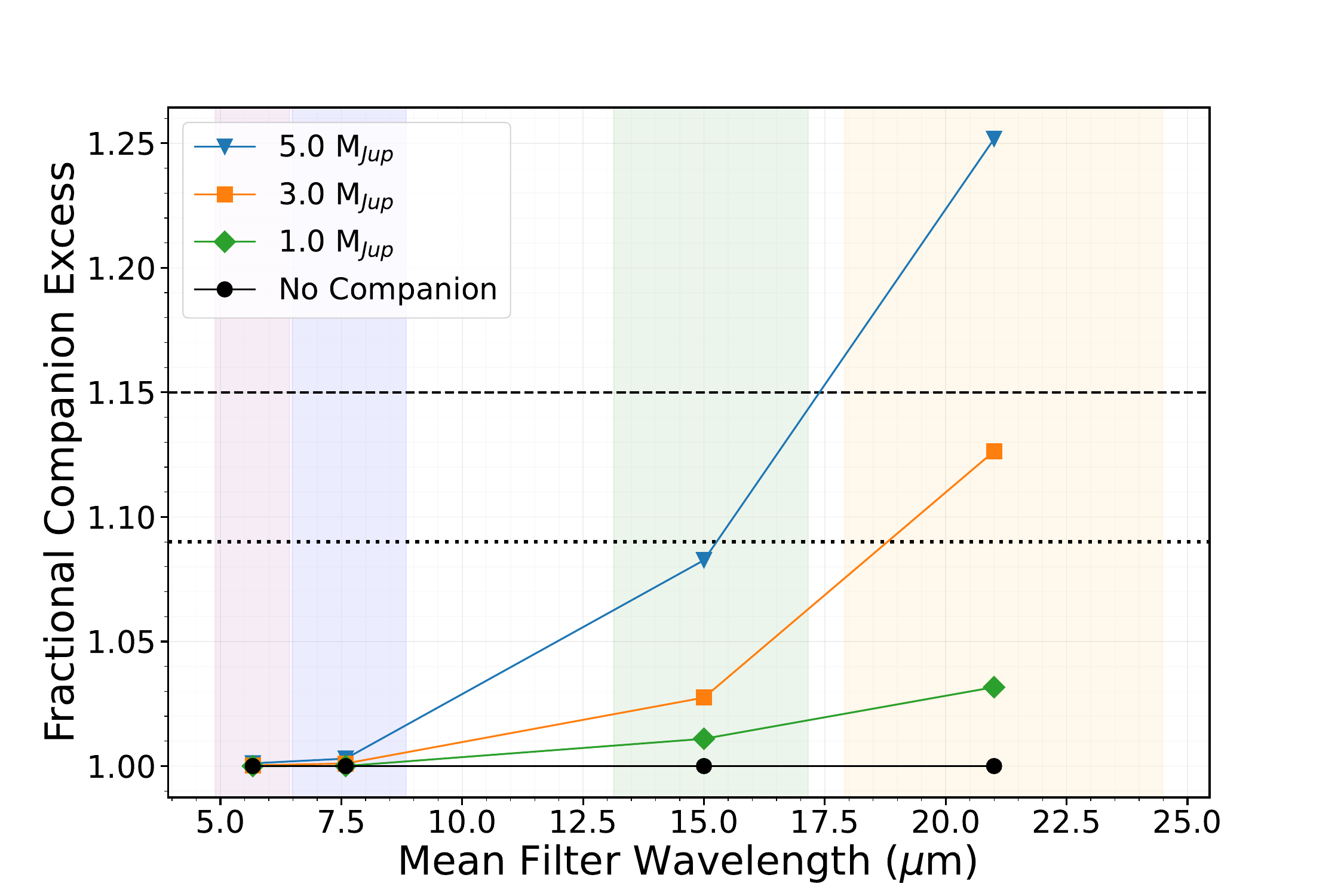} 
    \caption{The expected flux for \WD\ with varying mass companions, normalized to the expected flux of \WD\ with no companions, for each of the four MIRI filters used. From left to right, these filters are F560W (purple), F770W (blue), F1500W (green), and F2100W (yellow). With current absolute flux calibration reported to be 3$\%$ or better in all filters, the dotted and dashed horizontal lines denote a 3$\sigma$ and 5$\sigma$ confidence, respectively. Companions above 2\Mjup\ were modeled using cloudless Sonora-Bobcat grid models \citep{SonoraBob}, and companions below 2\Mjup were modeled using cloudless Bex models \citep{linder19}.}
    \label{fig:unresolvedDetection}
\end{center}
\end{figure}

To detect unresolved planets, we now focus on the 21 \micron\ filter as the white dwarf flux is lower than in the 15 \micron\ filter. We compared the photometry of the white dwarf against predicted spectral energy distributions (SEDs) of the photosphere, allowing us to search for infrared excesses in the longer wavelength bands. By modeling planet flux and adding that to the predicted flux of \WD\ with no companion, we find that planets close to $\sim$5~\Mjup\ would show a strong excess in the 21 \micron\ filter, while planets close to $\sim$3~\Mjup\ would still exceed a 3$\sigma$ significant excess (Figure \ref{fig:unresolvedDetection}). These limits will continue to improve as MIRIs absolute flux calibration improves.

JWST targeted \WD\ with the mid-infrared instrument (MIRI) in imaging mode with four different broadband filters: F560W, F770W, F1500W, F2100W (Tab.~\ref{tab:wdobsparams}). The object was targeted by JWST on 2022 October 13 and the data are located on MAST: \dataset[https://doi.org/10.17909/kj1r-9e95]{https://doi.org/10.17909/kj1r-9e95}.  Each image, grouped by filter, was composed of many exposures dithered using the Cycling dither pattern with the FASTR1 readout pattern. The exposure times and observing parameters are shown in Table~\ref{tab:wdobsparams}. Build 9.3 of the JWST Calibration pipeline was used to process the data starting from the uncal files. Each image was processed through stage one and two of the imaging pipeline using mostly default parameters as defined in the parameter reference files, only setting the jump detection threshold to 5$\sigma$. After the stage two pipeline was run, a mean background image was created for each filter and subtracted from each cal file. Lastly, the level three imaging pipeline was run on the background subtracted files and combined with the resample kernel set to `gaussian', the weight type set to `exptime', and doubling the outlier detection scale parameter to `1.0 0.8', but using the parameter reference files in CRDS for all other parameter settings. These background subtracted cal and i2d files were used for all subsequent analyses.

\begin{deluxetable}{crcccr}
\label{tab:wdobsparams}
\caption{Observing parameters of \WD. NGroups is the number of groups up the ramp per integration, and NInts is the number of integrations taken for each exposure.} 
\tablehead{
\colhead{Filter} & \colhead{$\lambda_0$} & \colhead{NGroups} & \colhead{NInts} & \colhead{Dithers} & \colhead{Exp time} \\ [-0.2cm] 
\colhead{} & \colhead{($\mu$m)} & \colhead{} & \colhead{} & \colhead{} & \colhead{(s)} 
}
\startdata
F560W  & 5.635  & 27  & 2  & 4  & 610 \\
F770W  & 7.639  & 10  & 2  & 4  & 233 \\
F1500W & 15.064 & 90  & 5  & 12 & 15118 \\
F2100W & 20.795 & 33  & 8  & 8  & 6016 \\
\enddata
\end{deluxetable}

\begin{figure}[h]
\begin{center}
    \includegraphics[width=\columnwidth]{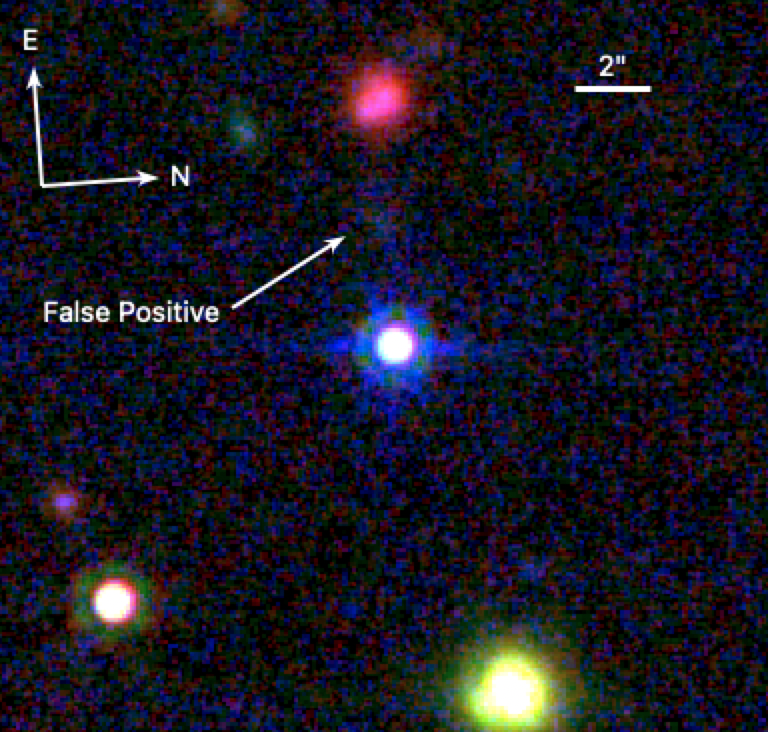} 
    \caption{Color image of \WD\ created by combining the F770W, F1500W, and F2100W filters. Visible directly above the white dwarf is a false positive candidate, eliminated due to its extended shape and incorrect color.}
    \label{fig:colorImage}
\end{center}
\end{figure}

Figure \ref{fig:colorImage} shows a color image of \WD, created by stacking the F770W (blue), F1500W (green), and F2100W (red) exposures. The white dwarf is in the center of the image, surrounded by background sources that are both unresolved and extended. A faint false positive, the closest source to \WD, is excluded as it is not consistent with a planetary SED and the FWHM along the sources major axis is twice the PSF FWHM for the F1500W filter (and is hence resolved). We assume point sources will have a FWHM $\pm$20\% from the nominal PSF.

\section{White Dwarf Modeling and Photometry}

In order to determine if \WD\ is orbited by unresolved planets or dust, we must first accurately model the white dwarf spectral energy distribution based on available photometry.  We then remove the white dwarf flux at the infrared wavelengths to see if there is any contribution from a much cooler counterpart. We first gather and fit against observed UV/Vis/NIR/Mid-IR photometry. For UV/Vis we convolve the HST CALSPEC spectrum of \WD\ that should have $\sim$1\% absolute flux accuracy \citep{bohlin01} with the GALEX FUV/NUV and Sloan $griz$ filter profiles via the Spanish Virtual Observatory SPECPHOT tool\footnote{http://svo2.cab.inta-csic.es/theory/specphot/}. For the NIR we use the 2MASS Point Source Catalog, and for the mid-IR we use a combination of ALLWISE W1, W2, and Spitzer SEIP flux values for IRAC2 and IRAC4 \citep{wright10,mainzer14,seip}. Table \ref{tag:photparams} gives the measured photometry for \WD. We then utilize the grid\footnote{https://www.astro.umontreal.ca/\~bergeron/CoolingModels/} of publicly available cooling models for hydrogen dominated atmosphere white dwarfs \citep{bergeron95,holberg06,kowalski06,tremblay11,bedard20} to minimize a chi-square metric assuming the Gaia DR3 distance \citep[$d$=22.4~pc]{gaia,gaiadr3} and estimate the 95\% confidence interval for \Teff\ and log~g. We find, from the photometric fitting, \Teff=17840$\pm$100~K and log~g=8.01$\pm$0.01.

\begin{deluxetable}{ccDD}
\tablehead{
\colhead{Filter} & \colhead{Central Wavelength} & \multicolumn2c{F$_\nu$} & \multicolumn2c{$\sigma_\nu$} \\ [-0.2cm]
\colhead{} & \colhead{$\mu$m} & \multicolumn2c{(mJy)} & \multicolumn2c{(mJy)} 
} 
\label{tag:photparams}
\decimals
\caption{Photometry of \WD}
\startdata
GALEX FUV & 0.1530 & 39.0 &  0.4 \\
GALEX NUV & 0.2353 & 50.0 &  0.5 \\
Johnson B & 0.4384 & 33.0 &  1.2 \\
Johnson V & 0.5480 & 28.2 &  1.7 \\
Sloan g & 0.4747 & 32.3 &  0.3 \\
Sloan r & 0.6188 & 23.8 &  0.2 \\
Sloan i & 0.7544 & 18.2 &  0.2 \\
Sloan z & 0.8791 & 13.8 &  0.1 \\
2MASS J & 1.2391 & 8.4 &  0.2 \\
2MASS H & 1.6487 & 5.0 &  0.2 \\
2MASS K & 2.1634 & 2.9 &  0.1 \\
WISE 1 & 3.4655 & 1.37 &  0.03 \\
WISE 2 & 4.6443 & 0.76 &  0.02 \\
IRAC 2 & 4.5024 & 0.79 &  0.02 \\
IRAC 4 & 7.8556 & 0.27 &  0.01 \\
F560W & 5.6173 & 0.49 &  0.02 \\
F770W & 7.6523 & 0.26 &  0.01 \\
F1500W & 15.1003 & 0.071 &  0.002 \\
F2100W & 20.8471 & 0.042 &  0.001 \\
\enddata

\end{deluxetable}

\begin{figure*}
\begin{center}
    \plotone{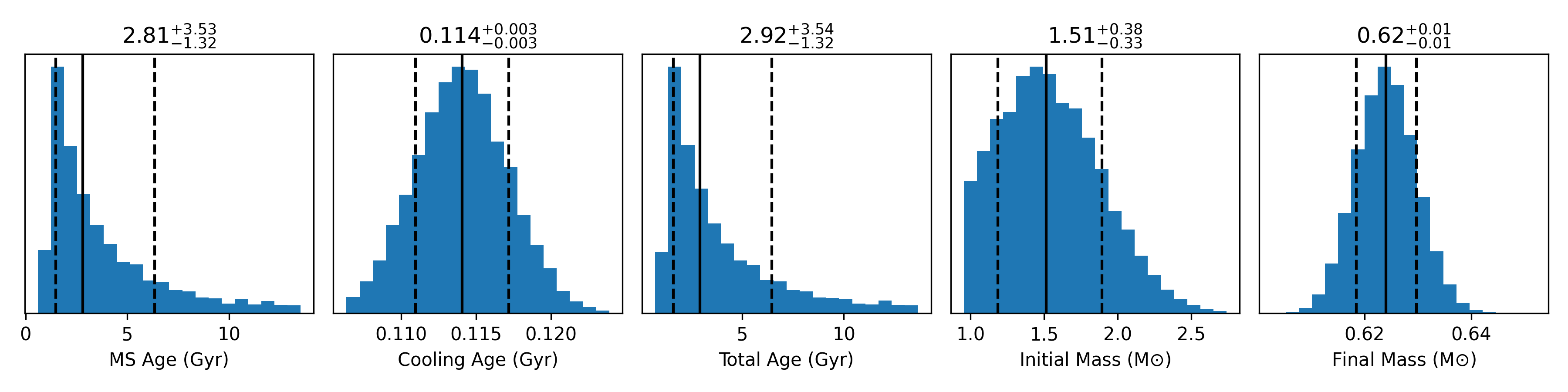}
    \caption{Probabilty distributions for the age, initial mass, and final mass of \WD\ derived using wdwarfdate \citep{kiman22}. The values \Teff=17753$\pm$72~K and log~g=8.01$\pm$0.01 were used for input values. The DA cooling model \citep{bedard20}, MIST based IFMR \citep{cummings18}, and [Fe/H]=0.0, v/vcrit=0.0 \citep{choi16, dotter16} were chosen for the model WD, model IFMR, and isochrone of the progenitor star respectively.}
    \label{fig:wddate_dist}
\end{center}
\end{figure*}

This is similar to inferred values from \citet{mccleery20}, who find \Teff=17876$\pm$188~K and log~g=8.01$\pm$0.01 and \citet{kilic20} who find \Teff=17753$\pm$72~K and log~g=8.06$\pm$0.004. All three of these results have consistent resulting temperatures and we agree with the surface gravity determined by \citet{mccleery20}. All three methods are similar: using a Gaia-derived parallax and photometry. While we use a larger number of filters, \citet{kilic20} used Pan-STARRs photometry and \citet{mccleery20} used Gaia DR2 photometry. The reported Pan-STARRS photometry appears to be fainter than would be expected from our Calspec spectrum, implying the possibility of slight saturation being present in the Pan-STARRS images, making this photometry unreliable for fitting purposes. Thus, we will assume a log~g=8.01 for calculations of the total age of the system. The mass and cooling age are then inferred from the cooling models and the WD mass-radius relationship using \Teff=17753$\pm$72~K and log~g=8.01$\pm$0.01 (Fig \ref{fig:wddate_dist}). This implies a mass of 0.62~\Msun\ and a white dwarf cooling age of 114 Myr.

\begin{figure}[h]
\begin{center}
    \includegraphics[width=\columnwidth]{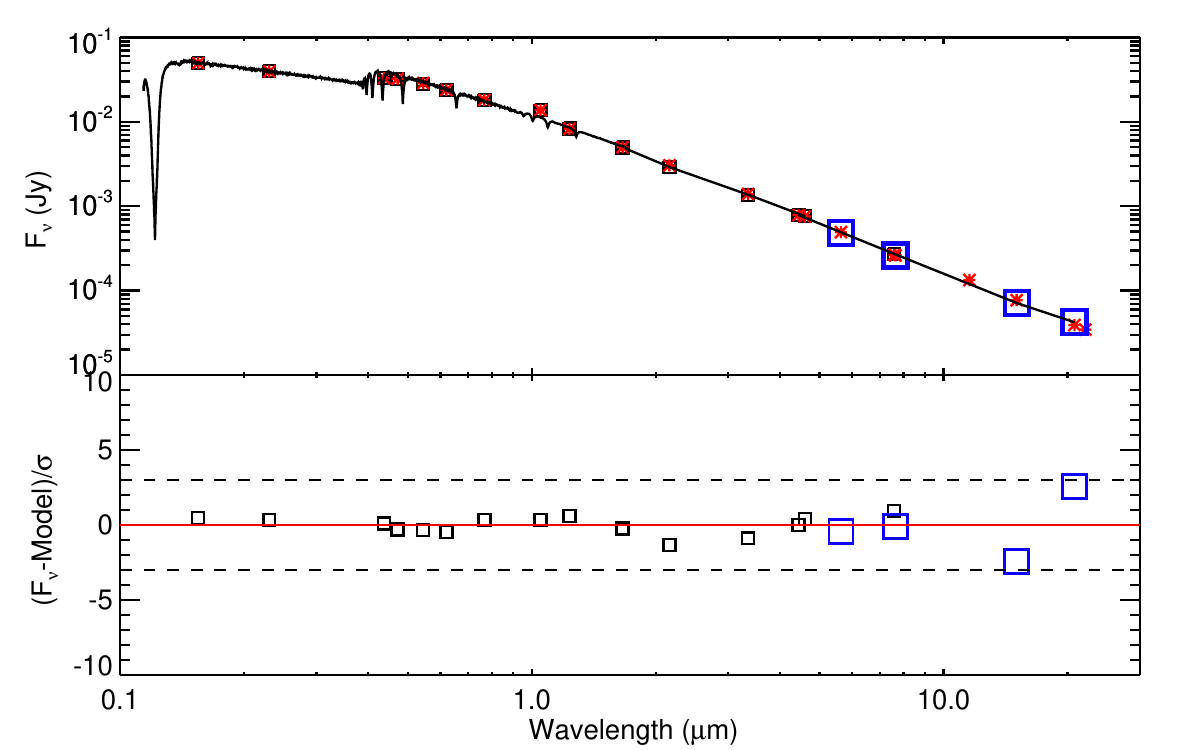}
    \caption{(above) The measured spectral energy distribution of \WD\ (black squares) compared to a DA cooling model with  \Teff=17480~K and \logg=8.01 (red squares). Overplotted is the CALSPEC spectrum for \WD\ (black line). The observed JWST MIRI photometry is shown with the blue squares and matches the expected photospheric flux to within the three times the absolute flux calibration uncertainties (dashed lines) of $\sim$3\%.} 
    \label{fig:sed}
\end{center}
\end{figure}

To measure the photometry of \WD\ in the four MIRI filters, we conducted aperture photometry using the recommended CRDS aperture radii and background apertures for each filter in the MIRI aperture correction file \citep{jdox16}. We chose photometric apertures that correspond to 80\% of the total flux of the target. We utilized the recommended color corrections for a Rayleigh-Jeans flux distribution, which correspond to 1.011, 1.021, 1.013, and 1.017 for F560W, F770W, F1500W, and F2100W respectively. Additionally, F1500W and F2100W have recently been reported to have experienced drops in sensitivity between commissioning and July 2023 of 3\% and 12\%, respectively\footnote{https://www.stsci.edu/contents/news/jwst/2023/miri-imager-reduced-count-rate.html}, with updates to the relevant calibration files now available. We have included the updated calibrations, as well as improvements to the absolute flux calibration for F560W and F770W, with expected uncertainties of 3\%. A comparison between the MIRI photometry of \WD\ and the expected photospheric photometry is given in Figure \ref{fig:sed}.

To predict the expected flux of the white dwarf for each filter, we interpolated the existing models to the reference wavelengths of the four MIRI filters. The resulting MIRI photometry is listed in Table \ref{tag:photparams}. We found that the MIRI photometry is consistent with the predicted white dwarf fluxes, and thus there is no evidence for a significant IR excess around \WD. 

\section{Results} \label{sec:results}
 In the following section, we demonstrate that our observations are sensitive to a variety of measurable signatures of planetary systems due to MIRI’s sensitivity, stable PSF and the absolute flux calibration. Cool giant planets can be detected either through infrared excess \citep[e.g.,][]{ignace01} or by direct detection via high contrast imaging \citep[e.g.,][]{burleigh02,debes05}. Additionally, warm dust within the tidal disruption radius of the WD can be detected as an infrared excess \citep{jura03}. Other more exotic signatures include infrared excess due to highly irradiated sub-Jovian planets \citep{sandhaus16} or the direct imaging of tidally heated exo-moons \citep{limbach13}.

\subsection{Limits to direct detection}
\label{sec:di}
In order to determine our sensitivity to resolved companions, the contrast between a circular aperture located at the center of the white dwarf (the central aperture) can be compared against multiple equally sized apertures located at a fixed distance away from the central aperture. The optimal aperture radius to maximize the SNR was determined to be 67\% \citep{taylor98,masci08} of the FWHM of MIRI's PSF\footnote{https://jwst-docs.stsci.edu/jwst-mid-infrared-instrument/miri-observing-modes/miri-imaging} for each filter. 

To obtain a meaningful contrast close to the WD and out to $\sim$10\arcsec, we constructed reference PSFs and subtracted them from the white dwarf in order to gain contrast interior to $\sim$1\arcsec\ in each filter. The MIRI PSF wings for the white dwarf at wavelengths shorter than 15 \micron\ are detected beyond the first Airy ring, which is field dependent and for the F560W filter also includes the cruciform feature \citep{wright23}.

\begin{figure}[h]
    \includegraphics[width=\columnwidth]{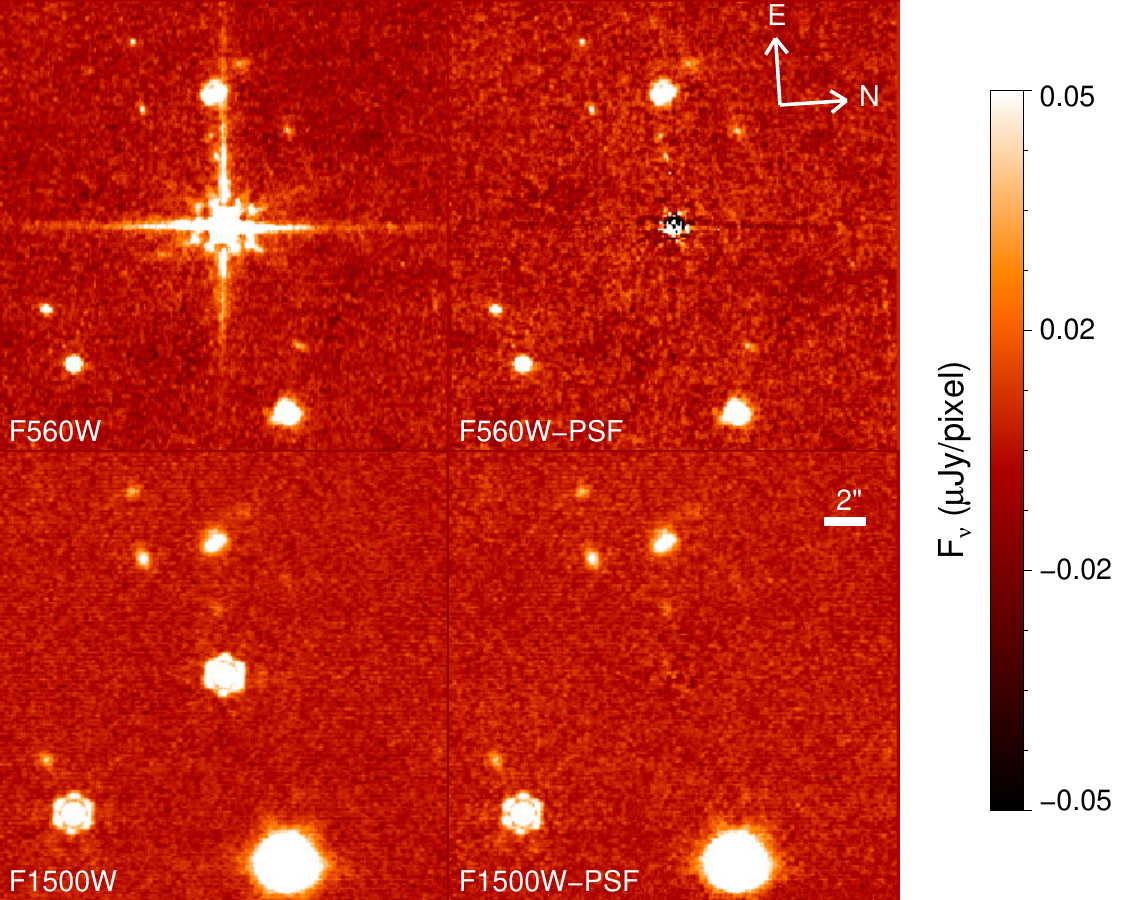} 
    \caption{(top left) F560W image of \WD. (top right) F560W with a reference PSF scaled and subtracted from the image. (bottom left) F1500W image of \WD. (bottom right) F1500W image with a reference PSF scaled and subtracted from the image. Our observations nearly hit the background sensitivity limit close to the star in both filters.}
    \label{fig:psfsub}
\end{figure}

For the F560W and F770W filters, we utilized the PSFs of two of our other targets: WD~2105-820 and WD~1620-391. We combined the two PSFs for each filter together and calculated offsets and scalings relative to \WD\ based on the measured stellar centroids and aperture photometry. However, the F1500W and F2100W exposures were much longer for \WD\ than for the other white dwarfs in our program, making it difficult to use our other targets for PSF subtraction. Instead, we used a bright star in the field of view to determine the shape of the PSF. Subtracting this PSF does not perform as well close to the star, but does perform better in the wings. Figure \ref{fig:psfsub} shows our PSF subtraction for F560W and F1500W.

 After calculating the flux in the central aperture using the non-PSF subtracted image, a ring of non-overlapping apertures is placed around the white dwarf on the PSF subtracted image. The aperture-to-aperture standard deviation was then measured for each ring, allowing us to calculate the 5$\sigma$ contrast for each radii as:
\begin{equation}
    \text{Contrast}(r) = \frac{5.0*\sigma(r)}{\text{sum}_{\text{WD}}}   
\end{equation}
where \(\text{sum}_{\text{WD}}\) is the total flux in the central aperture. Due to other objects in the immediate field around \WD, it was necessary to remove outlier apertures before taking their standard deviation. This process was automated based on the median absolute deviation (MAD) with a 5$\sigma$ rejection limit, such that an aperture would be rejected if:

\begin{equation}
   \frac{\text{sum}_{aper} - \text{median}_{ring}}{\text{MAD}} > 5.0
\end{equation}

\begin{figure}[h]
\vspace*{-1cm}
    \includegraphics[width=\columnwidth]{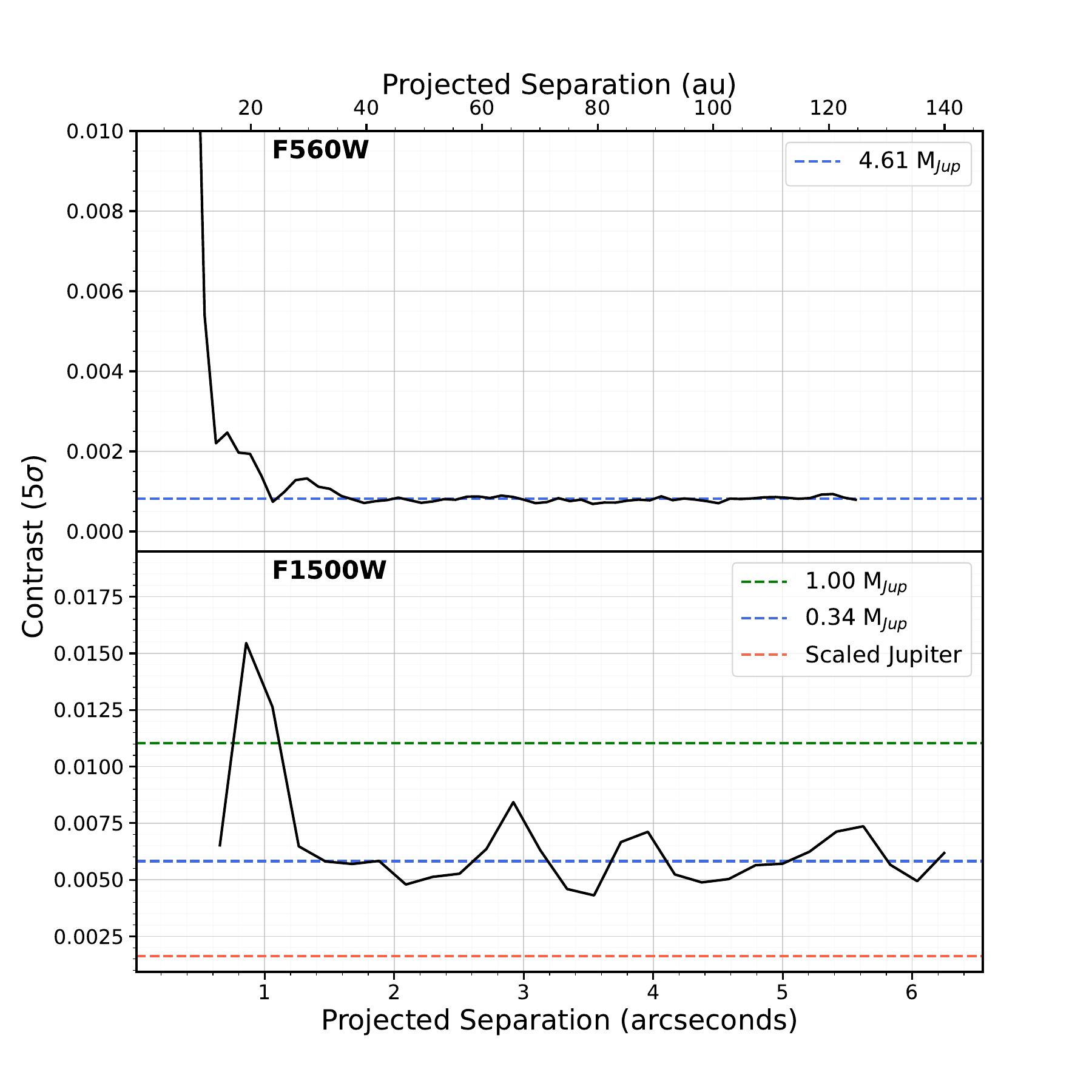} 
    \caption{This contrast curve reports the flux ratio of our central aperture relative to the standard deviation of the background at varying radial distances for two of our filters, F550W (top) and F1500W (bottom). The blue dashed lines are located at the median contrast level for each filter, and are reported as the associated limiting mass by interpolating the 3 Gyr Bex model. On the F1500W contrast curve, an additional line in green has been added to show the contrast level expected for a 1 Jupiter mass companion, also based on the 3 Gyr Bex model. Finally, the red dashed line shows the contrast level for our own Jupiter (4.6 Gyr) based on synthetic spectra generated using the Planetary Spectrum Generator \citep{villanueva18}, which has been scaled to the distance of \WD.}
    \label{fig:contrastCurve}
\end{figure}

However, apertures close to the white dwarf (within the first 2 rings) were not allowed to be rejected. This process consistently rejected apertures clustered near the false positive marked on Figure \ref{fig:colorImage}, as well as the suspected background galaxy above it, with few other apertures being rejected. The contrast was calculated using this method for multiple distances in order to create the 5$\sigma$ contrast curve shown in Figure \ref{fig:contrastCurve}.

A grid search of the PSF subtracted image was also performed, where an aperture was placed at each integer pixel within 2\farcs5 of the central source in order to probe for outlier sources. Of these apertures (also sized at 67\% the FWHM of MIRI's PSF $\approx$ 0\farcs32 = 2.97 pixels), none had a sum above 5$\sigma$ significance.

\begin{deluxetable*}{rclcl}
\label{tab:wdcparams}
\tablehead{
\colhead{Filter} & \colhead{Inner Contrast} & \colhead{Inner Radius} & \colhead{Median Contrast} & \colhead{Median Contrast Radius} \\ [-0.2cm]
}
\startdata
F560W  & 6.1$\times$10$^{-2}$ & 0\farcs277 (6.2 au) & 8.1$\times$10$^{-4}$ & 1\farcs06 (23.8 au) \\
F770W  & 2.2$\times$10$^{-2}$ & 0\farcs360 (8.1 au) & 2.5$\times$10$^{-3}$ & 1\farcs38 (30.9 au) \\
F1500W & 1.1$\times$10$^{-2}$ & 0\farcs654 (14.7 au) & 6.0$\times$10$^{-3}$ & 1\farcs26 (28.3 au) \\
F2100W & 6.8$\times$10$^{-2}$ & 0\farcs903 (20.2 au) & 4.6$\times$10$^{-2}$ & 1\farcs46 (32.8 au) \\
\enddata
\caption{Contrast values for each filter at the inner most radius, as well as the median contrast. The inner most radius and the median contrast radius (the radius at which the contrast hits the median contrast) are listed in both arcseconds and au.} 
\end{deluxetable*}

In order to convert the sensitivity of our observations given in Table \ref{tab:wdcparams} to planetary masses, we used cloudless Sonora-Bobcat grid models \citep{SonoraBob} above 2 \Mjup\ and cloudless Helios grid models \citep{linder19} below 2 \Mjup. We first calculated the proper \Teff, \logg\ combinations for a given age corresponding to various substellar companion masses from the Sonora-Bobcat grid. We then interpolated JWST fluxes from the grids of predicted fluxes for ages between 1.5-10~Gyr. For the Helios grid, fluxes for specific ages for 0.3, 0.5, 1, and 2 \Mjup\ were pre-calculated. We then converted the predicted fluxes or magnitudes to apparent fluxes accounting for the distance to \WD, 22.4~pc. 

\subsection{Analysis of Background Objects}
Since the contrast curve is consistent with hitting the limiting background sensitivity at radii beyond 1\farcs2, we also investigated the possible rate of contamination from both resolved and unresolved sources in the field. We conducted a search for significantly detected sources above the background with the DAOFIND equivalent IDL astrolib routine FIND.pro, using the following parameters for our F1500W image: We searched for objects with a peak flux of 10 nJy/pixel, default roundness and sharpness criterion, and an assumed PSF FWHM of 4.4~pixels (0\farcs48). We avoided the ``bonus region" in the upper left of the detector and masked out edges where only one or two dithers contributed to the image to cut down on spurious detected sources. This left a total detector area of 9237~arcsec$^2$.

Each detection was vetted to determine if it was consistent with a resolved or unresolved source by fitting the detection with a 2-D Gaussian. The core of the JWST PSF is reasonably approximated by a Gaussian out to $\sim$1-FWHM, and in general the morphology of extended sources is smooth enough to also be approximated by Gaussian shapes.

Spurious sources with excessively small FWHM values, offsets from flux peaks, or extremely large FWHM values were filtered out, leaving a total of 119 sources besides the WD present in F1500W. We determined a rough cutoff between resolved and unresolved sources by looking at the population of all detections and noting that the width of the distribution centered around 4.4~pixels (the FWHM of the PSF core) varied by $\sim$20\%. Nearly all of the proposed point sources below this cut-off (27) also had F560W detections which were visually inspected, with only 3 sources being resolved at the shorter, higher spatial resolution filter. We flagged those as extended and removed them as point sources, leaving 24 point source objects and 84 extended objects in the field.

\begin{figure}[ht]
    \includegraphics[width=\columnwidth]{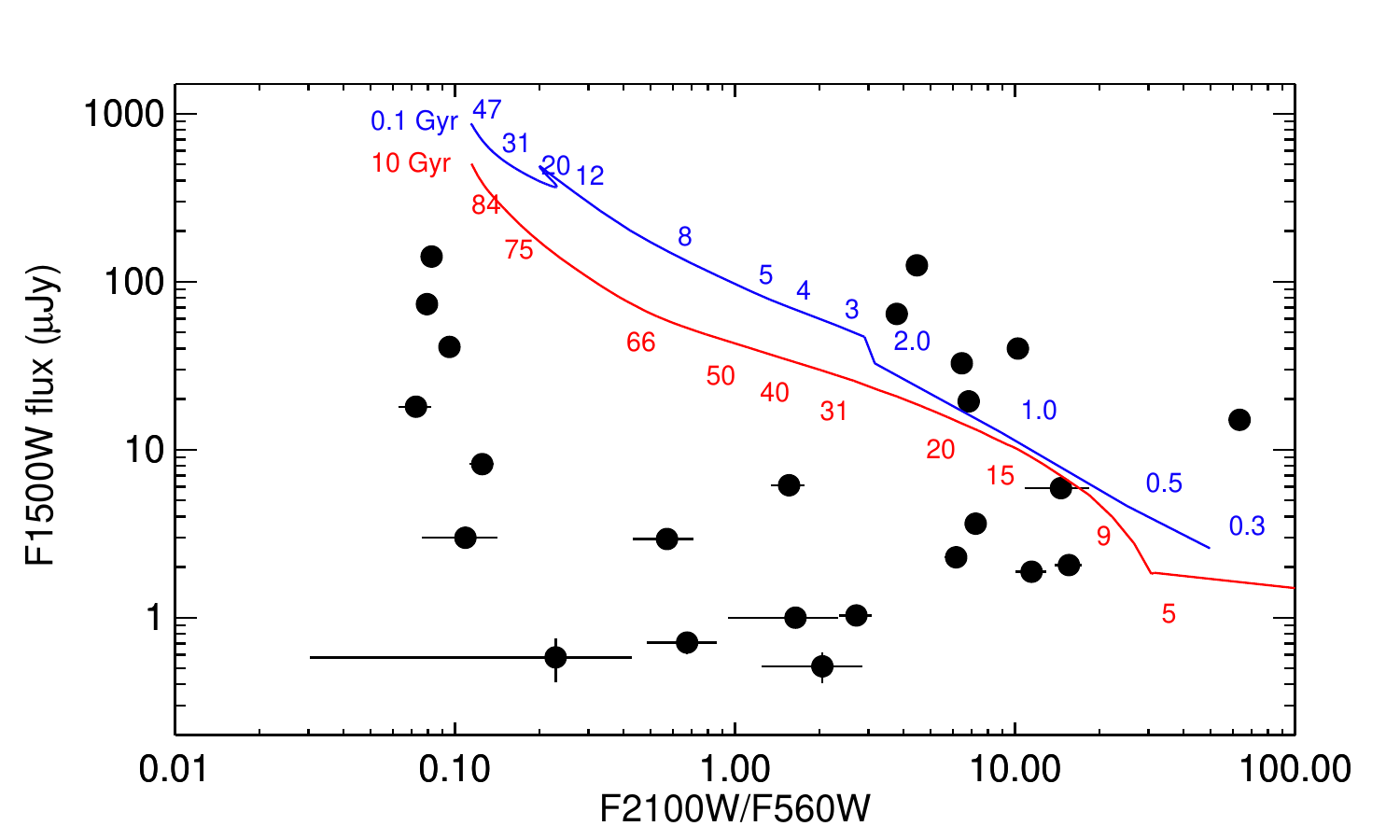}
    \caption{F1500W flux vs. F2100W/F560W flux ratio for all point sources detected in the \WD\ field. Overplotted is a 0.1~Gyr and 10~Gyr isochrone constructed from a combination of Sonora-Bobcat and BEX models, bracketing planets formed during post-main Sequence evolution to the upper limit to the total age uncertainty (6.5~Gyr, Figure \ref{fig:wddate_dist}) of \WD.}
    \label{fig:bgobj}
\end{figure}

Aperture photometry encompassing 60\% of a point source flux was measured for each detected object in the four filters. In most cases significant (SNR$\sim$5) detections were present in all filters. Figure \ref{fig:bgobj} shows the field point sources as a function of their F1500W flux and their F2100W/F560W colors. Unfortunately, exoplanet isochrones relevant to white dwarfs show that substellar objects overlap with background sources no matter what combination of colors is used for these four filters.

In the case of WD 2149+021, only a couple of objects are consistent, within the uncertainties, to predicted SEDs of cool Jupiter-mass planets. As mentioned above, however, there is still little observational constraint of what planets should look like beyond Jupiter and thus there is an inherent systematic uncertainty in our true sensitivity. Even so, it is clear that our observations at F1500W were more than sufficient to detect sub-Jovian mass planets, provided estimates of the F1500W flux are reasonably correct. One such object in Figure \ref{fig:bgobj}, which resides in between the two model isochrones and would be consistent with a 0.5-10~\Mjup\ companion, is also detected in an archival Spitzer IRAC2 image from 2009 (AOR: 35007232 PI: Burleigh). It is consistent with having zero proper motion and is thus firmly ruled out as a co-moving companion to \WD, which would have traveled $\sim$4\arcsec\ over the past 13~yr. A full common proper motion search for co-moving companions is beyond the scope of this paper. 

Given these results, we can estimate the probability that a background object will appear close enough to a white dwarf to potentially contaminate the star's flux or pose as a spurious planetary companion. First, we take the generic probability that any source will be present: in that case there are 0.01 sources per arcsec$^2$. Within the PSF FWHM at F1500W, the probability that there is a spurious IR excess due to a background object is just 0.8\%. For a search for planets within 100~au of the WD (4\farcs5), the probability of background contmination is 73\%. 

However, JWST is able to resolve most sources. In that case, the probability that a point source might be an interloper within 5\arcsec\ is 16\%, which can can be dependent on galactic latitude. Stellar objects typically have F1500W/F560W flux ratios of $<$0.3 under the assumption of a Rayleigh-Jeans flux distribution, which should not be the case for cooler planet mass candidates and extra-galactic objects. If we also filter-out objects with F1500W/F560W flux ratios $>$0.3, then the probability that a red point source will be within 5\arcsec\ to \WD\ is 11\%. Red point sources at separations of between 0\farcs5 and 1\farcs5 are highly likely to be physically bound to the WD, since the probability there is between 0.1-1.3\%. Larger imaging surveys should expect to find significant contamination from extended sources beyond a few arcseconds. Companions detected at smaller separations may be difficult to disentangle from resolved galaxies due to PSF subtraction residuals, although the overall probability that this will happen for the range of contrasts expected for planetary companions is fairly low. It is clear that any large surveys will be required to test for common proper motion, as is typical for direct imaging surveys at other wavelengths.

\subsection{Limits to unresolved companions}

After plotting the CALSPEC spectrum on top of our observed flux (Figure \ref{fig:sed}), we see that the observed JWST MIRI photometry matches the expected photospheric flux to within the absolute flux calibration uncertainties. The current accuracy of the MIRI flux calibration is 3\%. Assuming the 5-$\sigma$ upper limits of 15\% to the presence of unresolved companions, we can convert those upper limits to mass sensitivities to unresolved companions that are only emitting thermal emission with orbital radii of $<$0\farcs5 (11.2~au). Using the isochrones for a total WD age of 1.5, 3, 8~Gyr, corresponding to the youngest, most probable, and oldest total WD ages, we find limits of 2, 4, 7~\Mjup\ respectively. Even at ages as old at 10~Gyr (the maximum age of the theoretical isochrones), the mass limit is 8~\Mjup.

Interior to $\sim$0.3-1~au, the insolation from the host WD becomes important and contributes to the total luminosity of a companion, making it brighter than it otherwise would appear. Calculating an equilibrium temperature and a surface area under the assumption of blackbody emission allows us to place upper limits to the radius of objects that might be heated sufficiently close to the WD that might be detectable.

To estimate our radius limits, we calculated the maximum emitting area allowed by our 21~\micron\ 3$\sigma$ upper limits to the flux for WD 2149+021, following a similar procedure outlined in \citet{farihi14}. Using our flux limit ($F_{\mathrm {limit}}$) and distance to the WD ($d_{\mathrm {WD}}$) where we estimate the total emitting area ($A_{\mathrm {heated}}$) for blackbody emission $B(21~\micron, T_{\mathrm {eq}})$ with an equilibrium temperature ($T_{\mathrm {eq}}$):

\begin{equation}
\label{eq:aheating}
A_{\mathrm {heated}}=\frac{F_{\mathrm {limit}}d_{WD}^2}{\pi B(21~\micron, T_{\mathrm {eq}})}    
\end{equation}

with $T_{\mathrm {eq}}$:

\begin{equation}
    T_{\mathrm {eq}}= T_{\mathrm{eff}}\left(1-\alpha\right)^{0.25}\sqrt{\frac{R_{\mathrm {WD}}}{2~a}}
\end{equation}
where $\alpha$=0.3 is the planetary albedo and $a$ is the semi-major axis of the planet. We set R$_{\mathrm {WD}}=0.0127~R_{\odot}$. We assume a circular orbit. True planets' spectral energy distributions are modified from a pure blackbody by either their surface or atmospheric composition (e.g. Figure 6 from \citealt{limbach22}) and thus our true sensitivity is dependent on the properties of a given planet. We find that irradiated planets will be detectable if they range from 1-12~R$_\oplus$\ if they have orbital separations between the Roche radius and 0.3~au respectively. We can estimate the mass from the limiting radii, assuming the approximate relationship of 0.39~\Mjup$\times(R_{pl}/12.1~R_\oplus)^{1.8}$ \citep{bashi17}.

\begin{figure*}[ht]
    \includegraphics[width=\textwidth]{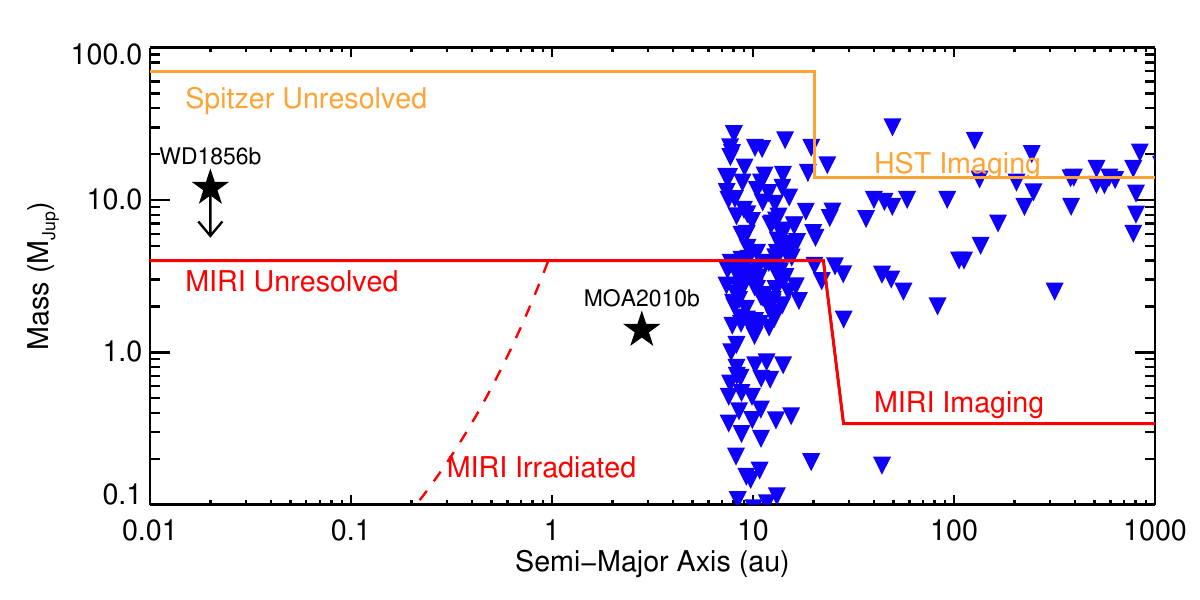}
    \caption{JWST/MIRI mass sensitivity for \WD. The solid red line denotes our mass sensitivity for \WD\ from unresolved IR-excesses and directly imaged companions. The dashed line represents IR-excess mass sensitivity assuming that close-in planets are irradiated by the WD. The orange lines are the IR-excess limits from previous Spitzer and HST direct imaging observations of other WDs. Symbols in blue represent planets with initial semi-major axes $>$3~au, the approximate orbital radius where planets are expected to survive post-main-sequence evolution \citep{mustill12}. Planet Semi-major axes are expanded to account for adiabatic mass loss without tidal effects. Our observations would be sensitive to 55\% of these known planet companions, if they were present in orbit around \WD. We also overplot the two known WD exoplanet candidates with semi-major axes $<$ 10~au. Exoplanet data provided by \dataset[https://doi.org/10.26133/NEA12]{https://doi.org/10.26133/NEA12}.}
    \label{fig:fullsens}
\end{figure*}

Figure \ref{fig:fullsens} shows a combination of the unresolved irradiated, unresolved thermally emitting, and direct imaging sensitivity limits from our MIRI observations as a function of orbital radius. We compare this to the confirmed exoplanet population known as of 1 Sep 2023, but with a modified semi-major axis for each planet. If every planet were to survive post-main sequence evolution and its orbit evolved adiabatically, its final semi-major axis would be a factor of 2.4 larger due to the mass loss of the host star as shown on our plot. For planets beyond $\sim$3~au, tidal effects from the red giant and asymptotic giant phases evolution are not strong enough to plunge the planet into the star. 

JWST observations can directly test the minimum survivable mass and semi-major axis of planetary companions in post-main sequence systems. As with post-common envelope binaries that have undergone common envelope evolution, we expect that some fraction of planets within 3~au to survive and migrate to small semi-major axes, where they are then irradiated by the white dwarf. Planets that survive this process move outwards and become amenable to direct imaging with JWST. Gaia and JWST will likely be sensitive to a wide range of planets down to $\sim$1~\Mjup\ that orbit between 1-10~au, though we expect most inner planets to be destroyed during post-main sequence evolution. For \WD, we would be sensitive to nearly 50\% of all currently known exoplanets with current orbital semi-major axes $>$3~au. In particular, we also show that our WD direct imaging is significantly more sensitive to intermediate and widely separated Jovian and sub-Jovian analogs than current direct imaging surveys in the NIR with ground-based AO.

\subsection{Limits to unresolved cold dust}

Most WDs that accrete rocky material do not show evidence for thermal emission from the dust that may be present. The current picture of this accretion requires rocky bodies that tidally disrupt, with constituent disruption products ground down into small dust which sublimates into the gas that eventually accretes onto the surface of the white dwarf. If there is sufficient material an optically thick disk could form, creating warm dust near the sublimation radius of dust within a few WD radii \citep{jura03}. This picture is consistent with the roughly 1-4\% of metal-polluted WDs that show significant Mid-IR emission at or within the tidal disruption radius \citep{debes11,barber14,rochetto15}. If the accretion is dominated by the Poynting-Robertson drag of dust, then certain dust accretion rates can account for sufficient mass in the disk to be observable \citep{rafikov11a,metzger12,kenyon17b}. Such disks are very easy to detect unless they are nearly edge-on. 

If the dust is not optically thick, the sublimation radius of dust for WDs like \WD\ moves outwards to roughly the Roche tidal disruption radius and can have much lower luminosities-- while an optically thick disk could reprocess up to a few percentage of the WD luminosity, optically thin dust akin to a debris disk can have orders of magnitude lower luminosity and remain undetectable to previous surveys of WDs.

However, colder or lower luminosity dust disks are amenable to detection with JWST. It is likely that there exists a larger number of low luminosity or cold disks around WDs with lower accretion rates. For example, recent work has shown that tidally disrupting rocky bodies do not form geometrically flat disks as originally envisioned by \citet{jura03} but rather can be vertically extended due to the lack of significant collisional damping \citep[hereafter K17][]{kenyon17a}. Such optically thin disks tend to be less massive while still consisting of a fairly large emitting area from small dust \citep{ballering22}. This would be consistent with existing circumstellar gas measurements that put the outer ranges of the gas at around the tidal disruption radius, since optically thin dust will sublimate further out \citep{manser,steele21}. So far only one accreting WD has been probed for additional cold dust beyond $\sim$10~\micron\ -- G29-38, which was probed both by ALMA and Herschel and showed no additional dust components \citep{farihi14}.

JWST photometry of nearby WDs at wavelengths beyond 15~\micron\ probes for very small amounts of warm and cool dust and start to constrain the details of dust accretion. \WD\ is sufficiently luminous to heat dust close to the tidal disruption radius, but significantly beyond that radius such that one would expect all planetesimals to be fully evaporated if they tidally disrupt \citep{steckloff21}. In fact, if the scenario of K17 is correct, that tidally disrupted bodies relax into quasi-circular swarms that collisionally evolve, JWST is sensitive to the predicted dust disks present for a large fraction of accreting white dwarfs. K17 predicts that steady-state dust disks formed from the regular injection of 1-100~km planetesimals result in a dust disk with an equilibrium mass of collision products of 10$^{14}$-10$^{20}$~g for accreting WDs.

In the case of WD~2149+021, the Ca abundance inferred by \citet{koester09} implies a mass accretion rate $\dot{M}$=2$\times10^7 g\ s^{-1}$ assuming a bulk earth composition. K17 found that the equilibrium mass ($M_{d,eq}$) for a collisional cascade at the tidal disruption radius ($a$) of a WD should be:

\begin{eqnarray}
\label{eqn:diskmass}
M_{d,eq} & \approx & 7\times10^{18}\ \mathrm{g} \left(\frac{\dot{M}}{10^{10} \mathrm{g\ s}^{-1}} \right)^\frac{1}{2} \left(\frac{0.6 M_\odot}{M_{\mathrm{WD}}}\right)^\frac{9}{20} \nonumber \\ 
& & \times \left(\frac{r_\mathrm{o}}{1 \mathrm{km}}\right)^{1.04}\left(\frac{\rho}{3.3 \mathrm{g\ cm^{-3}}}\right)^\frac{9}{10} \\
& & \times \left(\frac{0.01}{e}\right)^\frac{4}{5}\left(\frac{\Delta a}{0.2a}\right)^\frac{1}{2}\left(\frac{a}{R_{\odot}}\right)^\frac{43}{20} r_{\mathrm{o}}\ge1~\mathrm{km} \nonumber
\end{eqnarray}

where $r_{\mathrm o}$ is the characteristic size of the input bodies, $\rho$ is the average density, and $e$ is the eccentricity of the disk. The output gas accretion onto the WD is then equivalent to the influx of mass into the disk in a steady state \citep{kenyon17b}. For a collisional ring of bodies roughly at the tidal disruption radius $M_{d,eq}$=7$\times10^{17}$~g assuming r$_{\mathrm o}$=1~km. Interestingly, this also corresponds to just above the accretion rate predicted to show IR excess assuming late stage dust accretion from highly eccentric asteroids, an extension of the simulations done in K17 \citep{brouwers22}.

Similar to Equation \ref{eq:aheating}, we can estimate limits to unresolved dust emitting areas ($A_{\mathrm{dust}}$) where we calculate a new equilibrium temperature for dust grains instead of planetary surfaces.

We assume any dust present is primarily made up of silicates, and we use an empirically derived set of optical constants for galactic dust often used to model debris disks \citep{draine03}. We assume all the dust has a singular size $r_{g}$= 1\,\micron. We use the Planck-averaged estimates for an emission cross-section $Q_{abs}(T,a)$ given by \citet{draine03} for the relevant grain size and we assume a density ($\rho$) of 3.3 g cm$^{-3}$. Between dust temperatures of 100~K and 1300~K $Q_{abs}(T,a)$ ranges from $\sim$0.1-0.4 respectively, while $Q_{abs}(T,a)\sim0.1$ for 10~\micron\ grains. These quantities allow us to calculate an upper limit to the dust mass ($M_{\mathrm {dust}}$):

\begin{equation}
M_{\mathrm {dust}}=\frac{4}{3 Q_{abs}(T)}\rho A_{\mathrm{dust}}
\end{equation}

\begin{figure}
\includegraphics[width=\columnwidth]{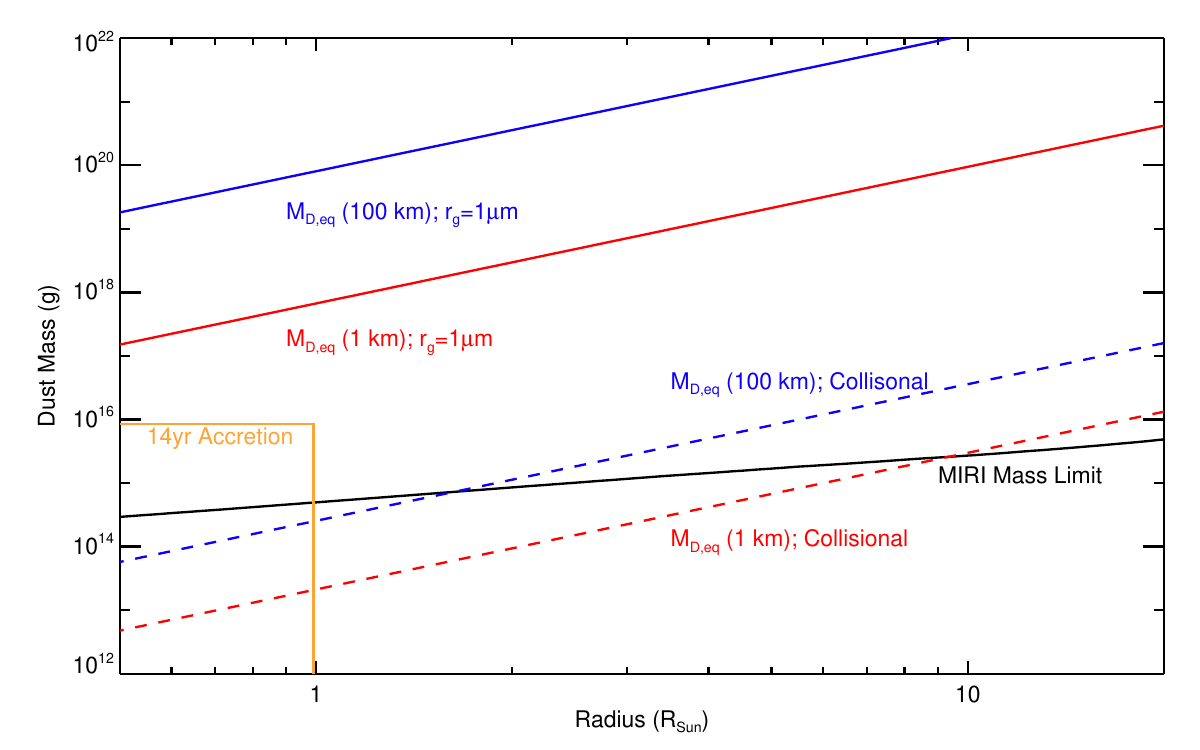}
\caption{Dust emission limits estimated for \WD. The solid black line shows the estimated MIRI dust mass limit assuming a mass in 1~\micron\ grains, derived from the F2100W upper limits to a flux excess. The yellow line corresponds to the total mass accreted by \WD\ in the time it's been known to possess Ca metal absorption lines. The solid blue and red lines represent the expected equilibrium disk mass of dust assuming a steady state collisional cascade (See Equation \ref{eqn:diskmass}) for typical planetesimal sizes of 100~km (blue) and 1~km (red). The dashed lines are the expected masses in 1~\micron\ grains only.}
\label{fig:dustlimits}
\end{figure}

Figure \ref{fig:dustlimits} shows our inferred dust mass upper limits with the above assumptions compared to a few different scenarios. First, we compare to the predicted equilibrium disk masses predicted by K17 (solid lines) in Equation \ref{eqn:diskmass} assuming disk radii from 1~R$_{WD}$ to several tens of R$_{\star}$, well outside the expected Roche radius ($\sim$R$_{\odot}$). 

If K17's predictions were correct, then we should have easily detected the dust with our MIRI observations. One caveat is that this assumes the dust is primarily in small grains, rather than locked up in larger bodies. To account for this effect we can modify the mass limits by the mass ratio of small dust grains to larger dust grains/bodies assuming a pure collisional size distribution, which is $\sim(\frac{r_{g,\mathrm max}}{r_{g,\mathrm min}})^{0.5}$, where we assume $r_{g,\mathrm min}$=1~\micron\ and $r_{g,\mathrm max}$ is either 1~km or 100~km. In this case, the total mass in 1~\micron\ grains is small enough to remain undetectable by MIRI exterior to 10 and 2 R$_\odot$ for cascades fed by 1 and 100~km bodies, respectively.

We also compare our expected mass limits if we assume that the total amount of mass in 1~\micron\ grains is equivalent to the last 14 years of accretion (orange line in Figure \ref{fig:dustlimits}). This is the total length of time that \WD\ has been observed to be actively accreting rocky material. Again, we would have easily detected that amount of dust, as our limits are a factor of $>$20 more sensitive interior to the Roche disruption radius.

Our MIRI observations demonstrate that a large survey of DAZ, DBZ, or DZ stars should be sensitive to very small amounts of dust, and that for accretion rates $>$10$^8$ g~s$^{-1}$, JWST is very sensitive to low luminosity dust excesses. 

\subsection{Limits to Tidally Heated ExoMoons}

We also consider the detectability of tidally heated exomoons that orbit giant planets around \WD\ and how that might impact candidate companions. Analogs to the Jovian moon Io might be tidally heated by very cool giant planets\citep{limbach13}. In this scenario, the moon has some fraction of its surface covered by volcanic activity at high T$_{\mathrm {eff}}$, with potentially enough short wavelength emission that it appreciably modifies the F2100W/F560W color of a planetary companion.

In Section \ref{sec:di}, we relied on the very red predicted colors of giant planets to help identify possible planet candidates. We show below that this might be complicated if large tidally heated moons are common around WDs. Predicted emission of tidally heated exomoons is presented in other works that consider both the possible average luminosity from tidal heating or fractional coverage by hot spots \citep{limbach13} with a longevity that in principle could last for billions of years \citep{roviranavarro21}, depending on the moon's size and orbital configuration. 

Recently, Io was observed with JWST/NIRSPEC and the mid-IR emission was consistent with a combination of temperatures and emitting areas ranging from T=1500~K and 2$\times$10$^{8}$~cm$^2$ to T=300~K and 1.6$\times10^{14}$~cm$^2$ \citep{depater23}. The combination of all the components creates a spectrum that peaks at 17~\micron, approximating a single blackbody spectrum with a T$_{\mathrm {eff}}\sim$320~K and emitting area with an equivalent area of 1.5$\times10^{14}$~cm$^2$. At 22.4~pc, Io would have a flux density in F1500W of $\sim$0.02~nJy, four orders of magnitude below our F1500W detection limits. At the given temperature and distance of \WD, detectable moons would need emitting areas 5 times larger than the radius of Io, and equivalent to a moon with R=1.36 R$_\oplus$. 

Since the population of exo-moons is poorly known, we take a simple approach that qualitatively illustrates the issue that a warm or hot exomoon poses: we calculate the emission of an Earth-radius moon with a fractional coverage of 1500~K hot spots relative to the \WD\ field point sources, and combine its mid-IR emission with two 3~Gyr planets with different masses, one with 2~\Mjup, and the other with 5~\Mjup. We then determine how much of the surface must be covered in hot spots to appreciably alter the color of these planets in the observed MIRI filters.

\begin{figure}
\includegraphics[width=\columnwidth]{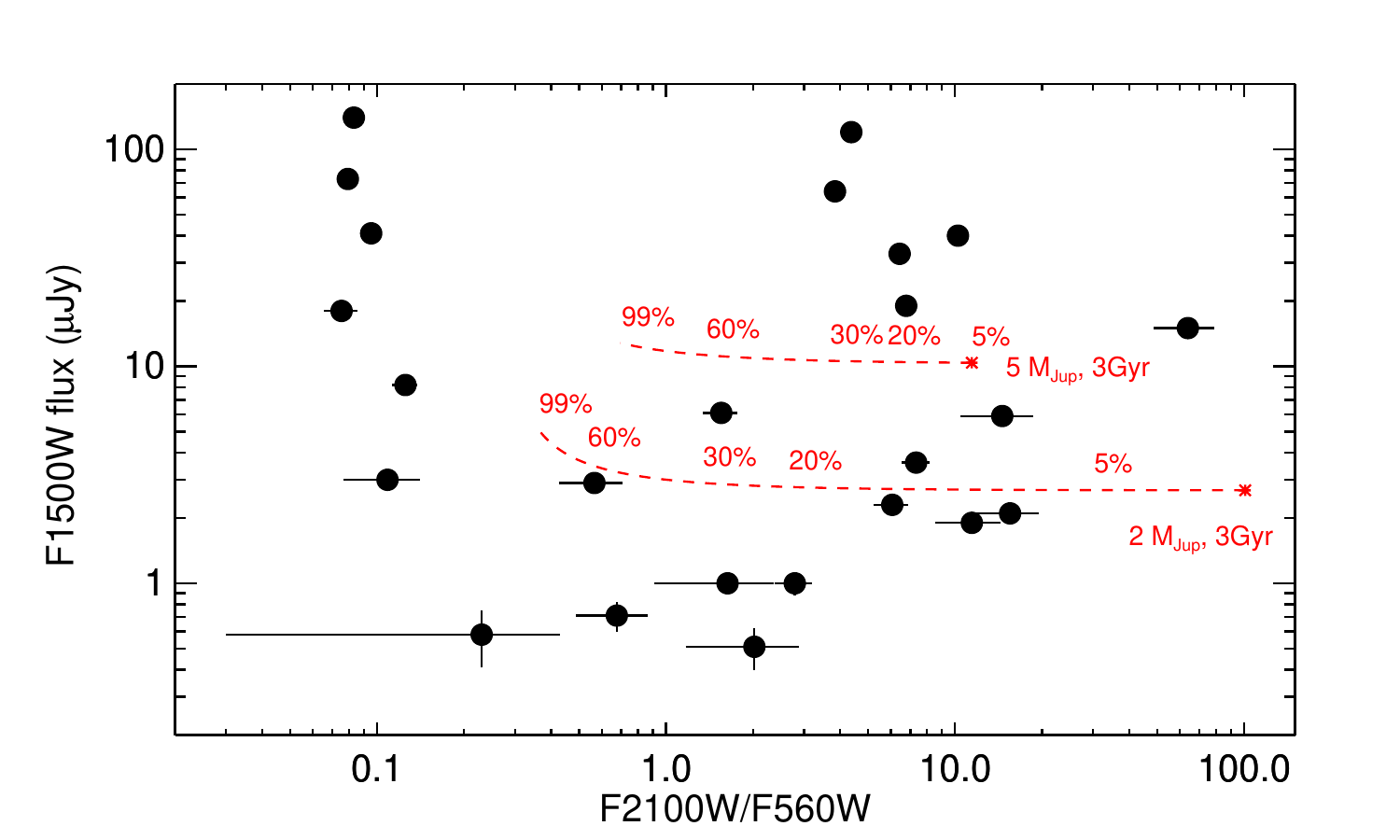}
\caption{F1500W flux vs. F2100W/F560W flux ratios for the field point sources in the field surrounding \WD. We compare these objects to the expected colors and fluxes for two planets (2~\Mjup\ and 5~\Mjup) that have 1~R$_\oplus$ tidally heated exo-Moons with varying coverage fractions of 1500~K hot spots. Significantly heated moons can potentially change the observed position of a Jovian planet candidate in this plot by having an excess of shorter wavelength emission.}
\end{figure}

The effect of a giant planet having a moon with significant short wavelength emission is to make the apparent flux ratio of F2100W/F560W less red, complicating the exclusion of sources with F560W detections. Whether large, tidally heated moons are common enough to significantly contaminate planetary searches around WDs is an open question, but the fact that these systems must be experiencing some dynamical perturbations suggest that it might be more frequent than around main sequence stars. Our deep MIRI observations, particularly at F1500W, are sensitive enough to start placing upper limits on the presence of large tidally heated moons in orbit around giant planets and reinforces the need for common proper motion tests to identify bound companions to WDs in the mid-IR.

\section{Conclusion}
We have presented the first multi-band photometry of \WD\ taken with the MIRI imager early in JWST's operational lifetime. To search for resolved sources we performed a grid search within 2\farcs5 of \WD, and to search for extremely widely separated candidates DAOFIND was used to detect round, resolved sources over the entire detector. To search for unresolved companions, we compared against available UV/Vis/NIR/Mid-IR photometry to determine if our observations show an infrared excess. This analysis did not identify any compelling candidates, placing limits on resolved companions of $\sim$0.34 \Mjup\ outwards of 1.263$''$ and $\sim$0.64 \Mjup\ at the innermost radius (0.654$''$). Limits on unresolved companions were found to be 2, 4, 7~\Mjup, corresponding to the youngest, most probable, and oldest total WD ages, respectively.

This is the most sensitive search for cold planets around white dwarfs ever conducted, reaching sub-Jupiter sensitivities for the first time. While recent discoveries of white dwarf exoplanets \citep{sigurdsson03,luhman11,gansicke19,vanderburg2020,gaia23} demonstrates that some planets are in close orbit and thus may survive their hosts' evolution, more observations are necessary to constrain the occurrence rate of these planets as well as their properties. It should be noted that improvements to the absolute flux calibration of the MIRI detector (we assume 3\%, but eventually it may reach 2\%) will improve our unresolved mass limits. Finally, the accuracy of planet mass sensitivities reported in this paper is limited by the accuracy of current exoplanet models, which do not focus on modeling at cold effective temperatures. Comparing against solar system objects such as Jupiter is a useful check for these planet models, but does not account for our system being $\sim$1.6 Gyrs younger.

\begin{acknowledgments}
Based on observations with the NASA/ESA/CSA James Webb Space Telescope obtained from the Mikulski Archive for Space Telescopes at the Space Telescope Science Institute, which is operated by the Association of Universities for Research in Astronomy, Incorporated, under NASA contract NAS5-03127. Support for Program number 1911 was provided through a grant from the STScI under NASA contract NAS5-03127. MK acknowledges support by the NSF under grant AST-2205736 and NASA under grant 80NSSC22K0479.

This publication makes use of data products from the Wide-field Infrared Survey Explorer, which is a joint project of the University of California, Los Angeles, and the Jet Propulsion Laboratory/California Institute of Technology, and NEOWISE, which is a project of the Jet Propulsion Laboratory/California Institute of Technology. WISE and NEOWISE are funded by the National Aeronautics and Space Administration. 
\end{acknowledgments}



%

\vspace{5mm}
\facilities{JWST, Spitzer, WISE, GALEX}





\bibliography{miriwd}
\bibliographystyle{aasjournal}



\end{document}